\documentclass[acmlarge,nonacm]{acmart}
\makeatletter
\newcommand{\myconfshort}{\acmConference@shortname}
\newcommand{\myconffull}{\acmConference@name}
\newcommand{\myconfdate}{\acmConference@date}
\newcommand{\myconfloc}{\acmConference@venue}
\AtBeginDocument{
  \fancypagestyle{firstpagestyle}{
    \fancyhead{}%
    \fancyfoot[C]{}%
  }
  \fancyhf{}
  \fancyhead[LO]{\@headfootfont\shorttitle}%
  \fancyhead[RE]{\@headfootfont\@shortauthors}%
  \fancyhead[LE]{\@headfootfont\footnotesize \myconfshort, \myconfdate, \myconfloc}%
  \fancyhead[RO]{\@headfootfont\footnotesize \myconfshort, \myconfdate, \myconfloc}%
  \fancyfoot[C]{}%
}
\makeatother
\acmBooktitle{\conffull\@ (\confshort), \confdate, \confloc}
\AtBeginDocument{\settopmatter{printacmref=true}}
\usepackage{quoting}
\AtBeginDocument{%
  }

\copyrightyear{2026}
\acmYear{2026}
\setcopyright{cc}
\setcctype{by}
\acmConference[FAccT '26]{The 2026 ACM Conference on Fairness, Accountability, and Transparency}{June 25--28, 2026}{Montreal, QC, Canada}
\acmBooktitle{The 2026 ACM Conference on Fairness, Accountability, and Transparency (FAccT '26), June 25--28, 2026, Montreal, QC, Canada}
\acmDOI{10.1145/3805689.3806498}
\acmISBN{979-8-4007-2596-8/2026/06}

\usepackage{tabularx}
\usepackage{rotating}
\usepackage{wasysym}
\usepackage{array}
\usepackage{colortbl}
\begin{document}
\title{Resume-ing Control: (Mis)Perceptions of Agency Around GenAI Use in Recruiting Workflows}

\author{Sajel Surati}
\affiliation{%
  \institution{New York University}
  \city{New York, New York}
  \country{USA}}
\email{sajel.s@nyu.edu}

\author{Rosanna Bellini}
\authornote{These authors contributed equally.}
\affiliation{%
  \institution{New York University}
  \city{New York, New York}
  \country{USA}}
\email{bellini@nyu.edu}

\author{Emily Black}
\authornotemark[1]
\affiliation{%
  \institution{New York University}
  \city{New York, New York}
  \country{USA}}
\email{emilyblack@nyu.edu}

\renewcommand{\shortauthors}{Surati, Bellini, and Black}
\renewcommand{\shorttitle}{Resume-ing Control:\\ (Mis)Perceptions of Agency Around GenAI Use in Recruiting Workflows}
\newcommand{\todo}[1] {{\color{blue}#1}}
\newcommand{\paragraphnew}[1]{\vspace*{3pt}\noindent\textbf{#1}\;}
\newcommand{\edit}[1]{{\color{black}#1}}

\newcommand{\thenote}{\thesection.\arabic{mynote}}

\newcommand{\rnote}[1]{\textcolor{teal}{\noindent $\ll${RB:} {\sf #1}$\gg$}}

\ccsdesc[300]{Social and professional topics~Computing / technology policy}

\keywords{Algorithmic Decision-making, Generative AI, Recruitment, Human-Computer Interaction, Behavioral Agency}

\received{24 January 2026}

\begin{abstract}
When generative AI (genAI) systems are used in high-stakes decision-making, its recommended role is to aid, rather than replace, human decision-making.  However, there is little empirical exploration of how professionals making high-stakes decisions, such as those related to employment, perceive their agency and level of control when working with genAI systems. Through interviews with 22 recruiting professionals, we investigate how genAI subtly influences control over everyday workflows and even individual hiring decisions. Our findings highlight a pressing conflict: while recruiters believe they have final authority across the recruiting pipeline, genAI has become an invisible architect that shapes the foundational building blocks of information used for evaluation, from defining a job to determining good interview performances. The decision of whether or not to adopt was also often outside recruiters' control, with many feeling compelled to adopt genAI due to calls to integrate AI from higher-ups in their business, to combat applicant use of AI, and the individual need to boost productivity. Despite a seemingly seismic shift in how recruiting happens, participants only reported marginal efficiency gains. Such gains came at the high cost of recruiter deskilling, a trend that jeopardizes the meaningful oversight of decision-making. We conclude by discussing the implications of such findings for responsible and perceptible genAI use in hiring contexts.

\end{abstract}

\maketitle

\section{Introduction}
\label{sec:intro}
Generative AI systems (genAI) are transforming the workplace, with 
studies reporting adoption rates 
up to 80\%~\cite{MIT, 
stanford_report}.
Recruitment is 
an exemplar of this shift: major platforms such as LinkedIn, 
alongside numerous startups, now integrate genAI 
across the recruitment funnel: from sourcing and screening to 
automating the interviewing process itself.

In high-stakes decision-making settings like hiring, scholars and policymakers emphasize the need for 
``AI-assisted'' decision-making, asserting that AI should support, not replace human judgment~\cite{lai2021towards,gdpr_art22}.
But how realistic is maintenance of human agency in practice? While prior work has both problematized and praised ``human in the loop'' or ``AI assisted'' decision-making through controlled experiments~\cite{algo_in_the_loop, algo_loop2, algo_loop3, hitl_1, hitl2}, there is little empirical work investigating how practitioners actually navigate incorporating AI–-especially generative AI---into their decision-making on the ground.
We use the field of recruitment and hiring 
as a case study to examine how agency is experienced and redistributed in AI-mediated hiring processes, both on the level of making individual hiring decisions, and systemically, regarding the adoption of or resistance to using genAI in their workflows. We focus primarily on the use of text-generative AI models with multimodal inputs, from ChatGPT to specialized tools that analyze audio and video from interviews.
Specifically, we investigate the following: 
\begin{itemize}
    \item \textbf{RQ1:} \textit{To what extent do recruiters feel they maintain agency in their decision-making? }
    \item \textbf{RQ2:} \textit{To what extent do recruiters perceive AI adoption as self-directed choice versus a necessity?}
    \item \textbf{RQ3:} \textit{How has shared human-AI decision-making changed the nature of recruiting decision-making processes, and what are the implications of these changes?}
\end{itemize}
We investigate these questions through interviews with 22 recruiters across eight industry sectors, identifying a widening gap between participants' perceived professional autonomy and the reality of AI-driven workflows. The mechanisms and implications of this phenomenon form our main contributions.

First, on an individual level (Section~\ref{sec:percieved}), we find that while recruiters claim ultimate authority---often asserting that \edit{either} AI does not make decisions or is relegated to low-level tasks---this presents an incomplete picture. 
Instead, our findings characterize genAI as an 
\textit{invisible architect}.
By defining job qualifications, creating interview rubrics, and summarizing candidate interactions, genAI structures the decision space and exerts a profound, unacknowledged influence on hiring---often before humans realize a decision is being made.
Second, on a systemic level (Section~\ref{sec:sys}), we find \edit{that} recruiters often lack agency regarding whether to include this invisible architect in their workflows.  
Many describe being boxed into using AI due to pressures to maximize personal efficiency, top-down mandates framing genAI as a business necessity, and an escalating arms race where candidate and competitor use of genAI requires automated countermeasures.

Finally, we highlight 
that the integration of genAI may lead to a subtle deterioration of human agency with three 
alarming repercussions (Section~\ref{sec:boiling}).
First, despite isolated instances of recruiters upskilling through AI use, we identify a broader risk of workforce deskilling in both domain expertise and the capacity to make deliberate, reasoned decisions about when and how to deploy AI, threatening professional oversight. This trend was particularly severe among junior recruiters.
Second, the AI arms race erodes recruiter trust in job-seekers' applications, pushing recruiters toward heuristic `vibe checks' that may intensify systemic bias, even as some recruiters perceive genAI as a more objective alternative. 
Overall, despite massive investment, according to our participants' reports, genAI has failed to transform applicant placement quality. Instead, the industry is inadvertently `boiling the ocean'---investing exhaustive effort for seemingly marginal gains while introducing new vulnerabilities like high turnover from poorly matched candidates.
Beyond these contributions, to map how genAI systems enter hiring workflows, we perform a functionality analysis of genAI hiring tools mentioned in interviews (Appendix~\ref{app:taxonomy}), and discuss how our findings inform paths towards reclaiming human agency in recruiting workflows through AI governance and future research in Section~\ref{discussion}. 

\section{Background and Related Work}
\label{sec:backlit}
\textit{Recruiting} refers to the process of identifying, attracting, and selecting individuals to fill specific job roles within a team or a company.
This may be performed by recruiters who are either in-house (often located \edit{with}in human resource departments), contracted out to an external agency, or performed by a current employee (a \textit{hiring manager}).
A recruitment pipeline consists of five chronological stages: job design, sourcing, screening, interviewing, and offering.
First, a job is designed or `themed,' whereby a person (typically a hiring manager) defines the responsibilities for a position.
Recruiters can either advertise the position to attract candidates (\textit{inbound}), or approach viable candidates (\textit{outbound}) with a proposition to apply.
Applicants are then screened, where their application materials are evaluated against the position's requirements and organizational factors; those who pass may be invited to an interview to evaluate their qualifications and `fit' across different teams or increasing staff seniority level.
A hiring manager may then select a candidate to extend a job offer.
Bias is well-documented in traditional hiring decisions, from résumé screening to culture-fit assessments \cite{kline2022systemic, bertrand2004emily, duncan2004never,cultralfit_1, cultralfit_2, cultralfit_3}. \edit{Indeed, algorithmic hiring tools are often marketed to reduce such bias~\cite{lessbias1,lessbias2,lessbias3,bogen2018help, manish}---yet as we discuss below, prior work suggests they introduce their own bias.}
\paragraphnew{Algorithmic Hiring Decisions.}
Our work studies the impacts of AI and genAI tools on an individual's perception of their agency in recruitment workflows. Prior literature highlights potential ethical concerns regarding AI in hiring, largely centered around the potential for discrimination. \edit{AI audits, following methodologies like those proposed by Costanza-Chock et al.~\cite{costanza2022audits}, enable investigators to find evidence of bias previously overlooked.} Several papers provide technical audits \edit{that} have quantified rates of racial and gender-based discrimination in both traditional AI~\cite{linkedin_talentsearch, christo_wilson} and generative AI~\cite{wilson1,glazko,bias_1, bias_2, bias_3, bias_4}, while sociotechnical audits have scrutinized claims made by companies developing recruitment tools\edit{~\cite{manish, sanchez2020does, sloane2022silicon}}. Closely related to our work, \edit{Li} et al.~\cite{min_paper} interviewed recruiting professionals regarding traditional AI in screening processes, though their focus remains on barriers to adoption. Similarly, Wilson et al.~\cite{wilson2} performed a user study on how humans are influenced by AI-generated candidate recommendations and demonstrate that humans are indeed influenced by AI-based recommendations. However, this study was not interview-based and did not include human resource professionals, \edit{instead focusing} on isolated recommendations rather than the broader recruitment pipeline. 
As prior work has focused on traditional AI or did not involve practitioners, this is, to our knowledge, the first study to interview recruiters directly about their use of \textit{generative} AI across the hiring pipeline.

\paragraphnew{Behavioral Agency in Human-Machine Interactions.}
Our work builds upon prior work regarding human agency in AI-assisted decision-making teams. 
In this work, we focus on \textit{behavioral agency} (derived from \textit{agere} \textit{``to set in motion, drive forward; to do, perform''}~\cite{agency_etym}), which focuses on an individual's or group's capacity to shape decision-making.
Behavioral agency has been explored across a wealth of complex sociotechnical systems~\cite{agency_survey, manish, user_reliance, sundar}, being so central to well-designed systems that developers have long been encouraged to \textit{``support internal locus of control''}~\cite{shneiderman}, despite the tension between human and machine agency~\cite{sundar}.
Human reliance on AI is driven by internal factors, such as trust and affinity for technology~\cite{relianceonAI, user_reliance}, alongside system traits, like perceived fairness or task simplicity \cite{explanations_overreliance}.
Thus, investigations into human-AI interactions suggest that decision-making is characterized by relational agency, where a human's capacity to act is constrained by an AI's architecture~\cite{maeda_when_2024, mertens_effectiveness_2022}.

GenAI complicates behavioral agency through introducing problems into the heart of decision-making.
For instance, LLMs can display \textit{``social sycophancy''}: human-like reasoning and mimic authoritative tones, whereby a model prioritizes saving face over providing critical or truthful feedback~\cite{malmqvist_sycophancy_2025, cheng_social_2025}.
Buçinca \cite{disruptive_design} and Lai \cite{lai2021towards} thus caution that the true locus of decision-making authority often resides in pre-computational stages: how an AI frames, filters, and prioritizes data before a practitioner intervenes.
\edit{Put simply, claims of choice can be done so within a \textit{``curated environment''} where the most cognitively taxing labor, the intentional, initial separation of data, has been performed by a technical blueprint~\cite{disruptive_design}.}
Prior work has highlighted further issues around navigating agency in human-AI collaboration \cite{position_paper}, including governance \cite{fin_govern}, high-stakes errors \cite{health_error1}, and deskilling \cite{deskilling_civics,deskilling_class,deskilling_health}. 
\edit{Given genAI's potential to affect a decision-maker's agency, it is imperative to learn how decision-makers navigate such effects. We do so in this work through interviewing recruiters about their use of AI in hiring workflows}.
\section{Methodological Approach}
Our study adopts a mixed methods approach, whereby we conducted a series of semi-structured interviews with 22 recruiters across the space of four months (Sept--Dec 2025) on their acceptance and use of such tools in their work practices (Section~\ref{sec:findings}).
We then paired this with a functionality analysis of the genAI hiring tools mentioned in their accounts (see Table~\ref{tab:taxonomy_vis},  Appendix \ref{app:taxonomy}).

\paragraphnew{Semi-structured interviews.}
Given the rapidly evolving nature of genAI integration into recruitment~\cite{linkedin_report1}, we chose to use semi-structured interviews for both epistemological and practical reasons.
As genAI is best characterized by idiosyncratic workflows that vary significantly between individual practitioners, we identified that a semi-structured format enabled the discovery of unanticipated insights: the `how' and `why' behind AI adoption.
In addition, recruiters, particularly in large organizational contexts often have to navigate a tension between company policy and practical `workarounds'~\cite{saxena_algorithmic_2024}, or unofficial tool use (e.g., Shadow AI~\cite{shadowAI}).
Thus, we hoped that the conversational 
atmosphere of a semi-structured interview could help build rapport with participants,
 leading to candid disclosures.

\paragraphnew{Recruitment Strategy.}
Participants were recruited using a multi-pronged sampling strategy.
First, we conducted purposive sampling via LinkedIn ($N$=19), targeting employees at organizations via the active job postings search feature to increase the likelihood of reaching potential interviewees who were engaged in hiring.
An active advert constitutes a listing being discoverable by verified LinkedIn users and remains posted on the site for less than six months\footnote{LinkedIn Help Answers report \textit{``Job posts will stay open until you close the job manually or until the job automatically closes after 6 months. You’ll continue to be billed until the job is closed.''}}.
We chose to prioritize identifying potential applicants via adverts over searching for specific people to mitigate the algorithmic bias that is baked into LinkedIn's recommender engines, which prioritize high-follower organizations and `popular' content tailored to a researcher's personal browsing habits, professional connections, or job history~\cite{linkedin_talentsearch}.

We sent an initial outreach message to 560 individuals based in the US who met the following inclusion criteria: (a) worked in human resources, recruitment, or run or directly play a significant role in a micro or small business, and (b) reported using genAI in their hiring processes of their current role, or a past role of no later than eight months ago.
We chose a purposive sampling approach to cover a range of different states, company sizes, and industries, which included: technology, manufacturing, biotechnology, healthcare, entertainment, publishing, marketing, and nonprofit sectors.
While a total of 41 potential participants expressed interest via this method (7.3\%), our yield was substantively lower at only 19 participants despite follow-ups (3.4\%).
To supplement this approach, we also used convenience sampling ($N$=1) by contacting professional acquaintances who had previously described using genAI in their recruitment workflows. 
Finally, the study also attempted to use referral-based (i.e., snowball) sampling ($N$=2) by encouraging participants to disseminate our study details within their professional networks, which yielded a low conversion rate.


\paragraphnew{Interview Process.}
Our interview protocol was designed around five topic areas: introductory conversation, grounding the context, identifying how genAI has been used in hiring, their perception of agency over choices regarding their work, and the future potential of AI.
For our second phase, we also leveraged the Critical Incident Technique~\cite{flanagan_critical_1954}, in which participants were encouraged to reconstruct their most recent hiring event to minimize recall bias and capture relevant data.
\edit{Additionally, we leveraged a normalization framing inspired by the results of Catherine et al. \cite{shadowAI}, which describe hesitancy to admit the use of AI tooling, for two questions pertaining to deskilling and the perceived benefits of AI in our protocol to gently challenge our participants' internal logic.}
Our interview protocol underwent three iterations and refinement.
We refined the protocol's timing and language through an internal trial with a member of the research team, followed by a pilot interview with a recruitment professional that revealed additional insights. 
Due to the richness of the pilot data, we chose to include their account in the final analysis ($P_{1}$).
Interested readers can find a copy of our interview protocol in Appendix \ref{app:protocol}.

Data were collected via synchronous, semi-structured interviews conducted over Zoom video-conferencing software. 
Each session lasted approximately 60 minutes (46--64 minutes) and was facilitated by a moderator and a designated notetaker to ensure that all data were captured effectively.
The research team also informed participants of future plans for the data, their timeline for concluding the interviews, and the option to receive a preprint of the research prior to publication for their personal use.
Data that were collected were automatically transcribed using Zoom auto-transcribe before being manually verified for spelling and turn-taking errors by the first author of this work for discrepancies.
Personal identifiable information, such as names, company organizations, and specific locations were replaced with placeholders to mitigate the unnecessary sharing of identifying data outside of collection.
De-identified transcripts were then stored and analyzed in \edit{a} secure Box folder only accessible to members of the research team.

\paragraphnew{Analysis.}
Our interview data were analyzed using a reflexive thematic analysis approach~\cite{braunclarke} using the qualitative analysis software \textit{Atlas.ti}~\cite{atlas_ti}.
Following familiarization by conducting a close reading of the transcripts, we used a hybrid approach of inductive and deductive coding.
We drafted an initial set of deductive codes ($N$=14) in accordance with our research questions (Section~\ref{sec:intro}), \edit{drawing from prior work on human agency in AI-assisted decisions \cite{agency_survey}}. These codes were then supplemented with a round of inductive open coding from two transcripts, resulting in 67 additional codes, for a total of 81 initial codes.
The research team then independently coded a subset of the transcripts to test the codebook’s applicability and refine definitions.
The codebook was refined further through two subsequent cycles with synchronous meetings for coordinating disagreements for a final codebook of 54 codes.
Inter-rater reliability of Krippendorff's Alpha were calculated as a mechanism for consensus-building, at 0.80 agreement (\textit{``Very Good''}~\cite{krippendorff}).
See Figure~\ref{fig:codebook} in Appendix \ref{app:codebook} for a copy of our final codebook.

To further investigate the unique tools mentioned by our interviews, we first confirmed that such software were regularly used by other industry professionals (i.e., non-bespoke) with popular aggregators~\cite{bestATS}. 
Then, we identified AI features across each system, then categorized them according to two qualities: what stage of the recruitment pipeline the tool is intended to be used for (\textit{recruitment phase}), and, what type of action being performed (\textit{functionality}).
Our results can be shown in Table~\ref{tab:taxonomy_vis} in Appendix \ref{app:taxonomy}.



\paragraphnew{Ethics and Limitations.}
This study \edit{was} reviewed and judged to be exempt by \edit{the New York University} 
Institutional Review Board (IRB).
Participants were sent an information sheet and a consent form prior to the interview, and written informed consent were taken prior to the interview with participants being asked to reconfirm their consent verbally at the start of the session.
Audio recording was voluntary and de-identified note taking was offered as an alternative.
Participants were provided a \$50 USD honorarium as remuneration for their time and contributions.

Given the potential for social desirability bias~\cite{braunclarke}, whereby participants could feel pressure to report policy-compliant or `ethical' use of genAI, we designed our protocol to utilize a normalization framing to reduce feelings of discomfort.
To further protect our participants from potential retaliation by employers, we chose to not report company names or sizes, report agnostic job titles, and report participant demographics in aggregate where possible \edit{(e.g. age), only including information willingly provided by participants}. \edit{Please refer to Table \ref{fig:participants} for participant details.}
We also report our findings as aggregated themes whereby idiosyncratic details (e.g., software, team structure) are withheld~\cite{bellini_sok_nodate}.

Our study also has the following limitations. 
As our participants report their genAI use retrospectively,
this may have limited their ability to accurately recall specific prompts or actions.
We also encountered a minority of participants ($N$=3) who omitted specific details of their work practice citing company policies. 
Finally, 
only 
interviewing participants in the US means that the international perspective is underexplored, which we look forward to \edit{investigating} in future work.


\begin{table*}[]
\small
\begin{tabularx}{.99\columnwidth}{llllll}
\toprule
    \textit{\textbf{ID}} & \textit{\textbf{Age}} & \textit{\textbf{Gender}} & \textit{\textbf{Most Recent Job Title}}     & \textit{\textbf{Exp.}} & \textit{\textbf{Tools Discussed}} \\ \midrule
$P_{1}$  & --                      & --                         & Chief Executive Officer                     & --                                      & ChatGPT, n8n                                  \\ \hline
$P_{2}$  & 35-44                      & Man                         & Sr. Recruitment Specialist                            & 17                                      & Internal system                                  \\ \hline
$P_{3}$  & 35-44                      & Man                         & Director of Recruitment               & 15                                      & BrightHire, Ashby, Gemini, Gem                               \\ \hline
$P_{4}$  & --                 & --                      & Talent Acquisition Specialist                              & --                                    & Indeed, ChatGPT                                  \\ \hline
$P_{5}$  & --                      & --                         & Director of Recruitment                      & --                                      & \begin{tabular}[l]{@{}l@{}}Metaview, Internal tool, Ashby, Granola, ChatGPT \end{tabular} 

                             \\ \hline
$P_{6}$  & --                      & --                         & Chief Operating Officer                     & --                                      & LinkedIn, ChatGPT                                \\ \hline
$P_{7}$  & 35-44                 & Woman                    & Director of Recruitment              & 22                                    & LinkedIn, ChatGPT                                  \\ \hline

$P_{8}$  & 35-44                      & Woman                         & Sr. Recruitment Specialist & 10                                      & Undisclosed AI agent service, Greenhouse, Mettl             \\ \hline
$P_{9}$  & 45-54                      & Man                         & Director of Recruitment        & 27                                      &         Workday, Gemini                          \\ \hline
$P_{10}$ & 35-44                      & Woman                         & Talent Acquisition Specialist                              & 13                                      & Paraform, LinkedIn, Ashby, Juicebox, Metaview                                  \\ \hline 
$P_{11}$ & 25-34                      & Woman                         & Sr. Human Resources Manager              & 8                                      & ChatGPT, LinkedIn, Zoom                                  \\ \hline
$P_{12}$ & 25-34                      & Man                         & Talent Acquisition Specialist               & 5                                      & Willo, Greenhouse, Crisp                                   \\ \hline
$P_{13}$ & --                      & --                         & Talent Acquisition Specialist                  & --                                      & ChatGPT, Gemini                                   \\ \hline
$P_{14}$ & 35-44                      & Woman                         & Talent Acquisition Specialist                              & 14                                      &  LinkedIn, Internal LLM, ChatGPT, Google Meet                                 \\ \hline
$P_{15}$ & 35-44                      & Woman                         & Sr. Recruitment Specialist                      & 15                                      &         ChatGPT, Greenhouse, LinkedIn                          \\ \hline
$P_{16}$ & 35-44                      & Woman                         & Talent Acquisition Specialist                  & 6                                      & ChatGPT, LinkedIn                                  \\ \hline
$P_{17}$ & 25-34                 & Man                      & Talent Acquisition Specialist                                   & 4                                     & ChatGPT, LinkedIn                                  \\ \hline
$P_{18}$ & --                      & --                         & Talent Acquisition Specialist                         & --                                      & ChatGPT, LinkedIn                                  \\ \hline
$P_{19}$ & 35-44                 & Woman                    & Talent Acquisition Specialist                  & 11                                    &             ChatGPT, LinkedIn, Workday   \\ \hline
$P_{20}$ & 35-44                      & Woman                         & Sr. Recruitment Specialist                  & 15                                      &     HireVue, iCIMS, Copilot                              \\ \hline
$P_{21}$ & 18-24                      & Woman                         & Talent Acquisition Specialist                      & 1                                      & Copilot, Internal system, LinkedIn                                  \\ \hline
$P_{22}$ & 18-24                 & Woman                    & Talent Acquisition Specialist                                   & 2                                     & Bullhorn, ChatGPT, LinkedIn                                  \\ \bottomrule
\end{tabularx}
\caption{Interview participant demographic information. ``Most Recent Job Title'' refers to a general class of job titles to retain anonymization. ``Tools Discussed'' only includes tools that were actively used by the participant, not merely mentioned in their interview.}
\Description{A table containing the participants' ID, age range, gender, most recent job title, years of experience, and the tools discussed in their interview. The age range of participants is from 18-24 to 45-54 and their years of experience range from one to 27. Their positions range from entry level to senior level. Additionally, 18 out of 22 participants discussed chatbots, like ChatGPT.}
\label{fig:participants}
\vspace{-0.5cm}
\end{table*}

\section{Findings}
\label{sec:findings}
We present our findings: how genAI subtly erodes human agency through acting as an \textit{invisible architect} in recruitment decision-making; how the very choice of whether to use genAI systems is not under recruiter control; and how this forced subversion of human agency may lead to bias, recruiter deskilling, and ultimately, little return on investment. 
\subsection{RQI: To what extent do recruiters feel they maintain agency in their decision-making?}
\label{sec:percieved}
Recruiters overwhelmingly reported feeling that they alone were in control of their hiring decisions. 
However, our interviews showed that genAI systems could potentially have more influence over these choices than many participants gave them credit for.
We identified that genAI systems were predominantly used to \textit{frame} decisions:
our interviewees shared many descriptions of genAI being used to 
both create assessment schemes of candidate quality, from job descriptions, interview questions, and rubrics. GenAI systems also \textit{processed} the information that went into presenting and evaluating candidates, such as summarizing resumes, interview performance, and more.
Thus, our analysis identified that genAI acted as an \textit{invisible architect}---an authority on organizing the framework in which people make decisions, and the information used in decisions, without the necessary visibility that could make participants aware of how their behavior was being shaped. As $P_3$ shared, 

\begin{quote}
    ``\textit{[AI] helps us to really identify the people we've been talking to, and boil down to the different patterns that exist... And then using that, it creates, like, an ideal profile, and then from there, you can create a job description. And then it also says, hey, based on this, here is an ideal interview plan. \textbf{And so it kind of helps you develop the entire operations that way, too, but also make decisions, because you are going to lean on it for some ... Some data, some information,} so on and so forth}.'' $(P_3)$
\end{quote}
 Here, $P_3$ acknowledges that by developing most of the materials used to assess candidates and processing all the information used to go into these assessments, genAI does actually influence the decision-making process itself because you must ``\textit{lean on it for... information}.'' 

\subsubsection{Recruiters feel in control and that AI only helps with low-level tasks}
\label{sec:low-level}
More broadly, we observed this dichotomy of perceived control and subtle AI influence through two co-occurring narratives. On the one hand, participants commonly reported feeling in control of hiring decisions
and chalked up the work of AI assistance, especially generative AI assistance, to low-level tasks. 
For example, regarding agency,  $P_{10}$ stated that ``\textit{I’m the deciding factor, right?}'', and $P_{9}$ stated that ``\textit{Bottom line, we don't use AI to make any, like, real decisions in hiring.}''  On the other hand, as we demonstrate in Section~\ref{sec:invisible_architect}, a considerable amount of decision-making power was passed on to AI systems via inherent trust of and deference to AI outputs, such as using AI-generated framing and AI-summarized information to make decisions. 

Part of participants' sense of control appeared to rest on framing AI as handling only low-level work: for example, $P_{18}$ stated that the internal generative AI tool in her office that she used to do the initial screening on candidates was ``\textit{basically like a filter. You know, it's filtering it so that it doesn't waste my time.}'' $P_{21}$ stated that the percentage match between a candidate's resume and job description that she prompts ChatGPT to come up with is ``\textit{A helpful tip}''. However, some of the descriptions of AI use had internal contradictions about its role and power: For example, $P_5$ said their AI interview tool ``\textit{Doesn't make decisions... It's completely just transcript only},'' despite noting that they used it to organize and highlight information from the interview: 

\begin{quote}
    ``\textit{If... I wanted to know... everything we talked about around compensation, for example, I can actually prompt the tool to tell me everything we talked about that involved a compensation}.''  $(P_5)$
\end{quote}

Another part of the gap between participants' assertions that AI does not make final decisions and their deference to it in practice may stem from participants' belief that AI should not make final decisions about candidates. 
As $P_{9}$ shared, ``\textit{I don't want to make any decisions with AI.}'' However, in practice, many recruiters ceded control by rarely questioning or verifying AI output.
As $P_{18}$ said about her AI filtering system discussed above, she ``\textit{[doesn’t] have to necessarily double-check}'' the choices it makes about who it filters out because she trusts the systems’ output.
This feeling of control even extended to recruiters using agentic AI systems that completely automate entire sections of the recruiting workflow, such as resume screening: As $P_8$ states, ``\textit{I can almost blindly move those 15\% \textnormal{[that an AI agent has approved autonomously]}... to talent interview, because I know that they have passed the bot}.'' Despite ``\textit{blindly}'' turning over decisions to the bot, $P_8$ said ``\textit{I still feel I'm in control... because... I am still the responsible one, so the bot is only gonna do what I tell the bot to do.}'' 



\subsubsection{AI as the invisible architect}
\label{sec:invisible_architect}
Despite participants' descriptions of how little AI and genAI tools influenced their decision process, our interviews demonstrated that genAI systems were invisible architects of the recruitment process. 
Everything from determining the relevant qualifications for a position, to generating interview questions and rubrics for assessing candidates, to summarizing and highlighting the most relevant pieces of information from an interview to make a final decision, were generated or summarized by an AI system. 
This fact, plus the common habit of deferring to AI output, means that while the `decision' about whether to select a candidate is often made by a human, 
genAI, subtly and incrementally, can significantly shape the hiring process.


For example, $P_9$ shared how genAI systems were essentially responsible for 
shaping the framework through which recruiters understand what skills the job requires and how they should evaluate and reason about applicants: 

\begin{quote}
    \textit{``Themeing, interview themes, ... like talking with a hiring manager about a job... We'll... transcribe that [conversation] and theme it and get all the information so then we can build the concepts for us, the questions for us, the rubrics based upon the job description, the panel interview questions ... We'll use AI to help, you know, organize the information and then generate things that are useful for us ... so Gemini will transcribe and theme, when we get it ... we don't have to do too much beyond that''} $(P_9)$
\end{quote}

Importantly, $P_9$ 
states here that 
the genAI outputs which frame decision-making don't need ``\textit{too much}'' done to them, i.e. little editing or oversight, and can be simply used in the recruiting process.  
Further, despite saying 
they ``\textit{use AI}'' to create ``\textit{things that help us assess the candidates in the process,}'' $P_9$ insists that ``\textit{In terms of, like, assessing applications, we don't use it as much as... some other organizations do}''---downplaying AI's influence, thus making it invisible.

Beyond AI designing decision-making frameworks, our interviews also showed that the processing of information relevant to assess candidate qualifications was often also left to AI.
Several participants 
  shared how they would use AI summaries to process and highlight information gleaned from candidate interviews, and often these summaries got moved straight along to higher ups. For example, $P_{12}$ stated:

\begin{quote}
    ``\textit{I use Crisp AI for note-taking in all of my interviews. It does a transcript and an audio recording and has an AI chatbot that you can ask `summarize all of their answers to the interview questions, and list all the questions they ask me verbatim' ... then I can stick that in there for the hiring manager.} $(P_{12})$''
\end{quote}

 Again, $P_{12}$ ``\textit{sticks}'' the AI summaries in the candidate\edit{'}s file for the hiring manager with seemingly little oversight. They went on to explain that while the transcripts exist as a ground truth for what was said during the interview, ``\textit{it's very uncommon that someone will actually go and review them.}'' $P_8$ seconded this practice, stating 

 \begin{quote}
     ``\textit{I kind of tell the bot what... to summarize out of this talk. So, for example, I would say, Summarize... the candidate's experience with enterprise-level clients and then the AI bot would understand that, and I can then fill that into the [candidate's] scorecard.}'' $(P_8)$
 \end{quote}
 
Adding to this pattern, $P_{10}$ states that thanks to the AI summarization, ``\textit{I don't have to do anything}'' to further process the interview. 
Evidenced by these examples, AI and genAI systems have huge influence over recruiting decision-making processes by designing assessment frameworks and processing relevant applicant information---but, recruiters do not recognize its influence, despite handing over control to AI. 
\subsection{RQII: To what extent do recruiters perceive AI adoption as self-directed choice versus a necessity?}
\label{sec:sys}
Our findings answer this question across three primary sectors of recruiting: the individual-level pressures to be more productive; the business-level pressures to survive; and, finally, how genAI is shaping the field towards pressures that resemble an arms race.

\subsubsection{Pressure to Optimize, Adapt, or Perish}
\label{sec:subfind:optimize}
Recruiters justified the adoption of genAI through a complex hierarchy of pressures, ranging from the pursuit of incremental \textit{``small wins''} ($P_{6}$) to a perceived existential necessity for professional survival. 
While many participants initially framed the use of genAI as a personal choice to improve their own efficiency, our data suggest such a choice was often a reaction to a saturated technological landscape where AI is perceived as ubiquitous and non-negotiable.
For many, the decision to integrate AI was a response to a psychological nudge. As $P_{16}$ reflected, the integration of genAI had become so normalized that it is often subconscious: \textit{``Yeah, probably, I use it [genAI] more than I realize... because I feel like AI is everywhere... if I'm feeling a little stuck... I'll use that.''}

This ``\textit{stuckness}'' ($P_{16}$) in creative work makes AI an easier path to choose, and one of least resistance, leading to an incremental surrender of agency over time. This pressure could also be further compounded by peer comparison; practitioners noted colleagues ``\textit{closing}'' cases faster, leading to a cost-benefit analysis of time. As $P_{18}$ noted: \textit{``If I can figure out how to use this ethically... It’s actually gonna save me a lot of time.''}
However, such a drive for efficiency in completing cases faster was not merely a personal preference but is 
often driven by
broader business pressures beyond themselves and their team. Many participants described situations where they framed genAI as a requirement (rather than a choice) for occupational survival.
As $P_{3}$ shared, 
\textit{``Every company's gonna be an AI company... if you're not using it, you're gonna die,''} with $P_{1}$ also echoing this sentiment, noting that avoiding AI could immediately make their company \textit{``less competitive.''}
Such pressures were also codified by top-down goals described by our participants, irrespective of the organization, that mold personal experimentation with genAI into compulsory corporate mandates. 
$P_{15}$ reflected on the shift toward embedding AI in annual objectives:


\begin{quote}
    \textit{``One of our corporate goals for next year is going to be utilizing AI to be more productive ... it's including all departments ... [leader] has assigned each of us ... to look at how we can be utilizing AI more.''}~($P_{15}$)
\end{quote}

As such, leadership also acts as a motivator for adoption, mandating that staff find ways to justify the technology within existing workflows. 
Such a top-down directive was often simplified in the daily experience of the recruiter, with $P_{20}$ succinctly noting: \textit{``The company is asking us to use it [genAI]''}. 
Nevertheless, such findings indicate a tension between the marketing promise of genAI---that it will seamlessly increase productivity---and the actual promised labor that organizations have to provide to implement it effectively.
Interviewees shared that in order to fill such corporate mandates (\textit{``use more AI,''} $P_{1}$), some organizations were investing heavily in onboarding and training, two areas that our practitioners mentioned were resource heavy.
As $P_{8}$ shares, ``\textit{success}'' with using such tools necessitated an adjustment of a rigorous training regime in onboarding new employees:

\begin{quote}
    \textit{``That's why we've decided to adjust and really train everybody that joins our team on this system, so that they know how to use the bots, because we have seen great success.''} ($P_{8}$)
\end{quote}

While we cannot comment if $P_{8}$'s adjustment is or is not positive, our interviewees pointed to the significant human labor required to \textit{``make genAI behave''} ($P_{13}$) in ways that were useful.
Furthermore, we identified that this could have an interesting ripple effect across team collaboration, where the perceived lack of being provided access to genAI was seen as putting their teams at a disadvantage.
$P_{1}$ reflected on a situation where employees on other teams were actively pushing for more access to genAI, even when the existing infrastructure around training or the justification for them gaining access to it was incomplete or unclear: \textit{``... they [a coworker] asked me last week, like, when is this [enterprise ChatGPT account] going live? Like, we need it... and I have to ask\edit{,} do you?''}

\subsubsection{Field-Driven Escalatory Pressures}
\label{sec:findings:arms}
An enormous factor that underlined all decisions around agency revealed a cycle of sociotechnical escalation whereby both job seekers and hiring practitioners felt compelled to use genAI based on the behavior of the other.
We characterized this by a feedback loop: \edit{as more candidates make use of genAI to optimize their applications,} thus increasing the volume of applications for recruiters to review, recruiters increasingly turn to `gut check' counter-measures to weed out percieved inorganic and AI-manipulated applications, as well as to reclaim their own agency in decision-making.

Interviewees cited their first adoption of genAI came in response to the perceived increased ease for applicants to apply for specific job roles.
Practitioners noted that genAI tooling enabled candidates to bypass a traditional `self-filtering' process, where a human may typically assess their own eligibility before applying.
This resulted in a surge of `noise' in the recruitment pipeline, as $P_{14}$ reflected when asked to reflect on genAI and applicant behavior: 

\begin{quote}
   \textit{``People are using AI to, you know, apply to some postings. Because [as] an individual [applicant], if he's reviewing it, he'll see that, okay, I might not be eligible for it, so no point of applying... It's completely irrelevant ... Why [would] a salesperson apply to a content editor position?!''} ($P_{14}$)
\end{quote}

Practitioners cited that this flood of irrelevant data was an incentive for them to rely on automated filtering systems to maintain a level of control. 
However, such a use of AI-to-AI systems creates a paradox as noted by many of our interviewees: as more candidates are perceived to use genAI to bypass filters, the more recruiters shared feeling the need to use genAI to filter them out.
This was best evidenced in how recruiters believed that candidates used AI to `mirror' job descriptions with such precision that it triggered suspicion rather than capturing attention. 
Thus, the perfect resume for the role, once the self-cited goal of recruitment, became a \textit{``yellow flag''} ($P_{22}$) for inauthenticity.
$P_{18}$ describes this reversal of trust, from being excited at a perfect candidate to being suspicious: 

\begin{quote}
    \textit{``There's no such thing as a perfect candidate, so when you see a perfect candidate, you go, okay, I'm gonna have to really dig into their experience ... they'll drop their resume into ... any type of AI tool and say, okay, here's the job description, like, basically fix my resume for it to present as this.''} ($P_{18}$)
\end{quote}

When the data points offered by a candidate felt \textit{``too polished''} ($P_{19}$), recruiters reported perceiving them as \textit{``a nice way to have something that sounds very polished ... but it's just so vapid''} ($P_{16}$) and lacking in genuine substance.
$P_{13}$ shared that this creates a state of noise and distrust where \textit{``exact resume matches''} were more likely to be treated with skepticism or suspicion because they often lacked the nuances around how they perceived applicants to pursue careers.
To reclaim such agency in a system that they perceive is being manipulated, recruiters described using ad-hoc detection methods to identify genAI involvement.
This was a shared strategy, 
where recruiters look for subtle linguistic or formatting cues that signal a perceived lack of human effort:

\begin{quote}
    \textit{``Whenever candidates have those bolded words, I'm like, okay, that's another, like, yellow flag... Is it a real candidate, like, tweaking their own resume? Is it something, like, an AI bot?''} ($P_{22}$)
\end{quote}

Thus, in reaction to the application overflow and the manipulation of previously reliable information, recruiters \edit{returned} 
to ``\textit{gut instinct}'' ($P_{20}$), potentially leading to \edit{the} 
dismissal of candidates \edit{due to} 
perceived genAI use. 


\subsection{RQIII: Implications of Changes to Decision-making Process and Boiling the Ocean}
\label{sec:boiling}
Our findings present a complex picture of genAI adoption and its impacts on recruiter agency in the realm of recruiting. 
While many of our participants described narratives that genAI provided a unique \textit{``secret thing''} ($P_7$) aiding efficiency and data-driven precision to find better candidates, the lived reality of practitioners suggests the presence of what we identify as a \textit{boiling the ocean} effect.
We observe this via three mechanisms. First, despite hopes for more objective and data-based decision-making, the use of genAI tools elevates the risk of bias through recruiters reacting to data overload by reclaiming agency through gut and culture-fit based hiring decisions. Second, despite the hopes of increasing recruiter skills through genAI use, we observed evidence of recruiter deskilling, threatening the maintenance of human agency in hiring in the long term.
Finally, while AI use is marketed as a time-saver, recruiter reliance on AI tools, despite their reportedly sub-par output, can lead to 
more work or lower quality outputs overall.
This, plus the increased use of AI by candidates increasing workload, suggests that AI use may not be delivering on its promises, as recruiters are ending up with lower-quality candidates, as evidenced by reported higher employee turnover.

\subsubsection{Gut- and Culture-fit based Decision-Making.}
\label{sec:findings:bias}
 Much prior work has noted that while AI often promises efficient and data-driven, and therefore unbiased, decision-making, automated systems can often further entrench bias~\cite{bias_1,bias_2,wilson1}. Our findings here add to this conversation by showing that, while participants did believe \edit{that} using AI tools would reduce bias, the use of genAI tools in recruitment introduced a less-discussed potential mechanism of bias: a retreat to gut-based decision-making that often relies heavily on assessing ``culture fit''~\cite{cultralfit_2}. 

\paragraphnew{Perception of Reduced Bias.}
Many participants clearly stated that they believed AI tools were a method of reducing bias in recruiting, as $P_9$ stated that he wanted to get ``\textit{past the biases of my recruiters}.'' He continued:

\begin{quote}
    ``\textit{If you take bias out, you may identify a bunch of candidates that we hadn't thought about previously, who may be able to deliver on that job... I'm really interested in how can I use AI to identify those people.}''~($P_9$)
\end{quote}
 Some participants also shared examples of when using AI systems helped them become less biased decision-makers. $P_1$ shared a story of how she did not want to hire a certain applicant because someone on her team already knew the person. But, ``\textit{when she was one of the, like, the shortlisted [applicants] by ChatGPT,}'' and $P_1$ thought her performance was ``\textit{very good}'' on take-home assessments, $P_1$ hired her---she concluded that it was ``\textit{very important}'' that ChatGPT was ``\textit{not biased}.'' While there certainly may be scenarios where the use of AI tools reduces bias, despite recruiters\edit{'} good intentions, we observed \edit{a different mechanism whereby }
  AI and genAI systems may be increasing bias in recruiting. 
 
 \paragraphnew{Gut- and Culture-fit based Decision-Making.}
 Due to the information overload that comes as a result of the onslaught of ever-rising numbers of AI-polished applications with little signal, and a need to remain useful as employees beyond AI tools, recruiters often expressed that one of the only meaningful choices they can make regarding candidates was based on gut instinct to assess what they believed AI could not replace: an understanding of business culture fit. 

Many recruiters spoke directly and indirectly about a retreat from data-driven decision-making.
Recruiters no longer trust digital artefacts (e.g., resumes, cover letters) that may have been fabricated with genAI (Section~\ref{sec:findings:arms}); many participants spoke to reverting back to \edit{their} \textit{``gut instinct''} ($P_{20}$) and a reliance on \textit{``vibe checks''} ($P_{8}$) as the final sources of truth in the recruiting pipeline. 
Several recruiters
spoke to their belief that what made them irreplaceable in a world of AI-based hiring was understanding business culture fit, which many have argued is a mechanism for unintentional bias in hiring~\cite{cultralfit_1, cultralfit_2, cultralfit_3}. While undoubtedly solely human-based hiring decisions also include reliance on culture fit, this increased emphasis on it as a reaction to AI may increase bias. 
$P_{18}$ stated that while she worried about ``\textit{At what point am I still going to be relevant, right? At what point are maybe interviews going to be taken over by an avatar completely, you know?}'' she pushed against the spectre of being replaceable by ``\textit{Learning how to ask more questions around, like, the human experience, which is what actually places people at the end of the day. And that's what AI can't do}.'' $P_{15}$ concurred:

\begin{quote}
    ``\textit{So much of it is, about our specific culture...I know sort of the microculture of this team, ... So I know, kind of, what's going to be a fit and what isn't, and AI doesn't know that}.'' ($P_{15}$)
\end{quote} 

Such a reliance on a recruiter's gut assessment of culture fit to combat both the fear of being replaced by AI and cut through perceived AI manipulation on the candidate side, may inadvertently reintroduce the same human biases that many participants believed genAI systems could screen out.
$P_{16}$ acknowledges this tension outright, highlighting the intrinsically messy reality of reclaiming agency in a new recruitment landscape: 

\begin{quote}
    \textit{``What's my feel on this candidate? Does it seem like they're aligned with what the manager's looking for? That's not a... like, AI's not gonna do that... So, there's so much room for bias, right? It's like, my bias is so imbued in all of our hiring and I have mixed feelings about that.''} $(P_{16})$ 
\end{quote}

\subsubsection{Deskilling.}
\label{sec:findings:deskilling}
Another common consequence of increased reliance on genAI that our interviews surfaced was evidence of recruiter deskilling around core elements of applicant evaluation. Importantly, however, there were also accounts of AI systems being used to acquire new skills. 

\paragraphnew{GenAI used to skill up.} 
\edit{Some participants hypothesized that although AI may lead to some currently used skills being underdeveloped, it may lead to the acquisition of others, for example ``\textit{[being] more creative}'' $(P_{10})$. $P_{12}$ describes this succinctly:

\begin{quote}
    ``\textit{I think maybe people that... just got into recruitment, like, this year, I doubt... [they] take as many notes... But they might have an entirely new book of skills at the same time that, you know, people who grew up before it [AI] didn't.}'' $(P_{12})$
\end{quote}}

\edit{This hypothesis was supported by several participants, who described} that genAI systems empowered them with efficiently learning about their hiring contexts. As {$P_{18}$} shared,\textit{``there's so many new technologies coming out''} relevant to jobs she places: 
\begin{quote}
    ``\textit{If I see, like, a job description, I also drop that into AI, ... okay, break this down in layman's terms, like, who am I looking for, candidate-wise? ... they'll really break down what is this technology.}'' ($P_{18}$)
\end{quote}

$P_{18}$ continued that this research support helped her sound more educated on the area she was hiring in, which helped forge stronger social connections with candidates, reflecting that genAI tools help to ``\textit{communicate things, because people [applicants] will hang up on you if ... you mispronounce technologies}.''
Some participants, such as $P_{15}$, even reported stories of directly using ChatGPT to skill up on the job: 
\begin{quote}
    \textit{``ChatGPT has helped me a ton with learning Excel ... it'll write me a super complicated formula, but then also tell me why it doesn't work. I am actually learning, even though it's doing a lot of the work for me. Now I can do a VLOOKUP by myself.''} $(P_{15})$
\end{quote}

\paragraphnew{Evidence of Deskilling.} Even when AI was used to empower recruiters with new information or save time, it often simultaneously deskilled workers.
Beyond simply atrophying skills, we also highlight deskilling as a lack of intentionality in the choice of tasks some participants assign to genAI, a skill in and of itself.

Sometimes deskilling was subtle, in a slight resistance to doing tasks which were previously a part of day-to-day job operations; other times, especially among more junior recruiters, the loss of agency seemed to result in a fear of or perceived inability to do parts of the job. 
Many participants 
described, alongside the ease of using AI, a reluctance or difficulty with performing tasks manually:  
as $P_1$ stated regarding reviewing resumes by hand,``\textit{I'm not going to read that. It's like you're reading a book.}'' $P_4$ found that a lot of resumes are ``\textit{Pretty wordy, or a lot of information},'' so it was ``\textit{nice to use that tool to summarize}.'' 
This reluctance persisted even if they acknowledged the product would be better or they would learn more. 
Both $P_{17}$ and $P_{18}$ talked about using ChatGPT to research their candidates' area of expertise and develop interview questions,
but shared they were conscious of depriving themselves of learning skills by using \edit{genAI systems }
to research: 

\begin{quote}
    ``\textit{I would have so much more knowledge if I sat down and had someone do a training on this technology instead of me just dropping it in [to AI] \edit{a}nd having it populate screening questions, because really, I'm just reading from a prompter, and at that point, you might as well put me out of a job, you know?}''~$(P_{18})$
\end{quote}

 

Other junior participants expressed fear of being unable to perform aspects of their job without AI assistance.
For example, $P_{21}$ shared that she was working on developing capabilities independent of ChatGPT, in case she ever lost access to her account:

\begin{quote}
    ``\textit{I'm just trying to get better about using it, and just, like, being more independent on it. ... I'm sure there'll become a day where ...I don't have access to it for X amount of time. I don't know what that would look like, but, like, I don't ever want to...be in a position where...I cannot do my job without ChatGPT.}'' $(P_{21})$
\end{quote}

$P_{13}$ stated that they like using genAI for aspects of communication they could not figure out on their own: ``\textit{I like using it for... sentences where I'm like, this makes no sense, and my brain is broken, but, like, I know I need to communicate this.}'' More senior recruiters also expressed frustrations with employee errors that were introduced by not checking over ChatGPT's outputs,  suggesting negligence to the point of deskilling: 
$P_9$ shared where his team directly used an unreviewed AI-generated candidate evaluation rubric and jeopardized the quality of the hiring process through inappropriate candidate rankings: 

\begin{quote}
    ``\textit{They're like... `should we have asked questions about it [the rubric item]?' Like, no, that shouldn't be in there. Right? And so, like, and that's the hard thing, like, you know, it wasn't the first question, it wasn't the last question, it was, you know, it was the rubric item number, you know, \edit{seven}. So, [they] didn't notice it\edit{...} so that's the kind of stuff that I really want them looking at.}'' $(P_9)$
\end{quote} 


\edit{We identified that when AI systems structure how decisions are made---such as by introducing changes to evaluation rubrics that participants “didn’t notice” $(P_9)$---they can reduce opportunities for deliberate human reflection on those decisions. This, in turn, can erode the professional skill of intentionally deciding when to delegate to AI versus make a decision independently.
Though some recruiters were very cognizant of what tasks they believed should and should not be allocated to genAI, others were not. 
For example, $P_{10}$ had a strong reaction to using AI in screening, saying that ``\textit{I don't trust it at all in that aspect,}'' while $P_{14}$ maintained that there should be ``\textit{no AI}'' in the interview process. 
On the other hand, when asked about whether there were tasks that should not be allocated to genAI, $P_{21}$ said ``\textit{I don't really think there's anything ... Yet, I'm sure there is something I probably shouldn't be doing with it, but I cannot think of it.}'' 
This absence of a clear boundary can make it easier for genAI to creep into more aspects of the hiring pipeline without clear reasoning.}

\edit{For participants who claimed productivity boosts (Section \ref{sec:hidden-labor}) from AI use, they had trouble justifying their use of genAI outside of those gains. This is clear in the thinking of $P_9$: ``\textit{I think any efficiencies we can create using AI, we try to use,}'' where efficiency comes first before the consequences of using it ``\textit{in all [of] the supporting functions around the picking, and then just a little bit in the picking}'' $(P_9)$. This \edit{gap in} 
clear intentionality in when to refer to AI in the decision-making process is echoed in $P_{18}$'s call for ``\textit{rules and boundaries}'' around AI-use.}

In the long term, such deskilling may erode the continued possibility of meaningful human agency over AI-assisted decisions \edit{and the ability to repair a pipeline that overrelies on AI}. 

\subsubsection{(Lack of) Return on Investment.}
\label{sec:findings:roi}
Participants cited improved efficiency or profitability as their primary justification for genAI adoption, particularly weighed against the risk of falling behind competitors who did.
This can be understood as a stand-in for a rough social calculation of return on investment (ROI); participants regularly cited the marketing promise of genAI---the ability to transform organizations and reap massive efficiency gains in daily work.
However, we identified that despite the investment in AI and the change in day-to-day operations, participants often described how genAI seemingly created as much labor as it purported to save, and ultimately did not result in higher quality candidates---ostensibly, the entire purpose of recruiting. 

\paragraphnew{Hidden Labor of AI.}
\label{sec:hidden-labor}
\edit{Many participants described the ways in which AI saved time in their daily work. These time-saving gains were non-negligible, where some claimed that AI saved ``\textit{60\% of your day}'' ($P_5$) and helped them ``\textit{massively}'' $(P_8)$.}
However, while participants shared that AI tools saved time in the short run, they also gave indications that the use of AI may create more work overall. 
As a basic matter, participants expressed frustration with the administrative and sometimes cognitive overhead required to make AI useful. 
For example,  $P_{21}$ shared that the level of precision needed for a successful ChatGPT output often rendered using the tool as \edit{an} inefficient use of time: \edit{\textit{``I'm just wasting my time... typing up this whole two-paragraph-long thing to then... I would have read that and figured it out myself at this point.''}} 


Further, participants shared that often, AI created worse outputs than they could on their own, but they would often use it anyway, potentially leading to low-quality work with implications downstream.
For example,  $P_{10}$ reflected that despite using genAI on the initial volume of applicants,
\textit{``manual Boolean searches''} were still required to ensure quality: \textit{``I still need to do it myself to produce the highest amount of quality [candidates]''} ($P_{10}$). However, crucially, many participants did not invest the manual work to fix AI outputs: when $P_{19}$ talked about her frustration with generating boolean searches, stating that sometimes when using ChatGPT she feels ``\textit{this isn’t working}.'' Yet when asked if she would try to fix if on her own or continue to use the tool, $P_{19}$ stated, ``\textit{some of both,}''  but often ``\textit{I just try it again later}’’ with the tool. $P_{18}$ shared, ``\textit{Sometimes it's helpful, sometimes it's not, but I find myself doing it [using genAI] all the time}.'' 

Beyond this loop of using AI tools even if they were unhelpful, the automated nature of these tools leads, in some cases, to professional embarrassment or \textit{``bitten-in-the-ass''} moments ($P_{17}$) where recruiters rely on saved templates or AI-written messages without a cursory review before posting or sending. Beyond the examples of this discussed above in deskilling (Section~\ref{sec:findings:deskilling}), whereby workers that move too fast lose the ability to catch these errors, leading to problems downstream, agency recruiters also spoke of their colleagues ``\textit{cutting corners}’’ with AI to meet their numbers in the short-term to the detriment of long-term gains:

\begin{quote}
\textit{``I've also seen, you know, recruiters cutting corners…they're like, oh, I can just send this candidate over, but it backfires, because you end up using Al editing their resume to make them look more presentable ... and ... when they show up in an interview, it doesn't present the same way, right?''} ($P_{18}$)
\end{quote}

\edit{As $P_{18}$ describes, though editing resumes with AI to increase perceived job fit may improve short-term  efficiency for recruiters incentivized to maximize candidate-job matches, it creates downstream inefficiencies for hiring managers, or can even lead to hiring an unqualified employee, potentially incurring high costs long-term.} 

Finally, as discussed in Section~\ref{sec:findings:arms}, candidate use of AI created an environment for many of our participants where they had to shift the focus of their agency from \textit{``finding talent''} to \textit{``detecting fraud''} ($P_{18}$). As $P_{7}$ shared, this can actually result in the advent of AI being a time suck overall, as recruiters have more and more irrelevant applications to scan: ``\textit{[it's] just such a time suck. Like, trying to investigate if someone is real}.’’

\paragraphnew{Low perceived ROI.} While participants were keen to emphasize that genAI may \textit{``help the day-to-day a little bit''} ($P_{18}$), this observation was coupled with the reflection that it had failed to deliver the perceived \textit{``insanely beneficial''} ($P_{18}$) transformation at the organizational or field 
level.
Many participants noted that the core metrics of what was considered success in recruiting (e.g., placement, sales) have not seen a corresponding leap forward:

\begin{quote}
    \textit{``No one's [case] numbers have gone through the roof, and... no company has, like, quadrupled in their placements and sales based off of it... I don't think it's moving things faster.''} ($P_{18}$)
\end{quote}

Overall, between recruiters being forced to use AI, leading them to eventually lose agency or cut corners,  and candidates using AI to generate falsely ``perfect'' profiles, in the end, recruiters painted a world where they are doing more work daily—going through more profiles—but ending up with worse outcomes overall, i.e. not finding high quality candidates. As $P_{14}$ shares, while ``\textit{every day these ChatGPTs, they are applying to 10 jobs}'' on applicants\edit{'} behalf, increasing the load of applications, $P_{18}$ explains that even with recruiter use of AI in response, this leads to more work in screening, and worse candidates overall:

\begin{quote}
    \textit{``I'm more screening more now, are you actually going to be able to show up and do the job and not just look like it on paper? ... That’s where we're having a lot of turnover, because they're using AI too much.''}~($P_{18}$)
\end{quote}

Thus, AI's efficiency gains are eroded. Instead of finding 
higher quality 
candidates---the goal of introducing AI---participants 
report a greater chance of hiring someone 
who must be replaced, “\textit{wasting everyone’s time}” ~($P_{18}$). 

\section{Discussion and Concluding Thoughts}
\label{discussion}
\edit{Our findings highlight that genAI systems have become ``invisible architects'' of the hiring process---they a large influence on hiring outcomes through generating decision-making frameworks (rubrics, job descriptions) and processing the data that goes into these frameworks (interviews, resumes), yet this influence is nearly imperceptible as each piece is seen as low-level work (Section~\ref{sec:invisible_architect}). Further, we find that while genAI offered participants some perceivable productivity benefits in tasks such as information gathering (Section~\ref{sec:percieved}), many reported struggling under the pressure to use such tools (Section~\ref{sec:sys}), even when participants deemed such use unsuitable.
GenAI's touted benefits may also be leading to a boiling the ocean effect (Section~\ref{sec:boiling}), whereby the efforts invested to make genAI `work' for their workflow was often described as contributing more work, unintentional de-skilling, or shaping subpar outputs that needed correcting. 
We believe these discoveries have implications for AI policy, organizational practice, and areas where further research is needed.}

\paragraphnew{Intentional Decision-Making Policies.}
Much law surrounding AI accountability 
focuses on decision-making outputs: 
for example, in the EU, the GDPR emphasizes the right not to be subject to \textit{decisions} \textit{``based solely on automated processing''}~\cite{xiang2024fairness, gdpr_art22}. 
However, our findings show that the introduction of genAI systems to high-stakes decision-making pipelines significantly broadens the sphere of influence that AI systems have over these decisions. 
Generative AI systems create the entire \textit{infrastructure} (see \cite{mertens_effectiveness_2022, saxena_algorithmic_2024}) within which that decision rests: organizational policy documents, rubrics, in addition to \textit{processing} the information that is considered, \edit{which may not be perceivable to its users}.
This creates a tension: recruiters, and other AI-assisted decision-makers, can claim that a final decision was human-authored, even as the decision space itself has been extensively structured by AI. 
As a result, legal protections 
around automated decisions may fail to capture the ways in which agency
is 
altered by genAI-assisted processes. 
Whether or not this is a problem is an open question,
however, our work shows that if regulation aims to limit the extent to which AI makes high-stakes decisions affecting individuals lives, our current regulatory frameworks may be insufficient. 

Relatedly, increased integration of human-AI decision-making may make it more difficult to comply with existing laws regulating high-stakes decisions in practice.
For example, in the U.S., Title VII prevents disparities in employment-related assessment outputs that cannot be justified by ``business necessity’’~\cite{eeoc_uniform_guidelines}. In addition, several new state laws requite mitigation of disparities of AI models used in high-stakes decisions such as hiring \cite{colorado, illinois}. Thus, AI systems to be used for hiring and other high-stakes decisions need to be tested for bias before deployment---but when human and AI decision-making is so intertwined, it is hard to develop tests that accurately portray how bias may arise during deployment.
More generally, the suite of current methodologies for discrimination testing, and other forms of legal compliance, are ill-suited to the level of human-AI collaboration happening in high-stakes decisions, creating a new set of problems for researchers to tackle.




\paragraphnew{Organizational Practices to Combat Deskilling.}
Many resources for employees in the workforce are focused on prompt engineering: how to prompt genAI better to produce improved results, \edit{promoting an implicit encouragement that genAI adoption is both legitimate and essential \cite{prompting1, prompting2}.}
This is a tempting response to presumed participant need; if recruiters just figure out how to use genAI, then this will result in time being saved (Section~\ref{sec:findings:arms}).
Our findings suggest that organizations should also focus on critical oversight of genAI use to prevent the rapid professional deskilling that may arise when recruiters \textit{``move too fast''} ($P_{18}$, Section~\ref{sec:findings:roi}) and question the role of genAI in occupational survival ($P_{3}$, Section~\ref{sec:subfind:optimize}).
\edit{To promote the} opportunity to learn and develop intentionality, \edit{despite} the allure of summarization and saving minutes but losing hours, we suggest that future onboarding should incorporate a form of intentional friction to any genAI use.
For example, pedagogical interventions that force recruiting professionals to articulate their professional logic behind any gut instinct ($P_{16}$, Section~\ref{sec:findings:bias}) could hold promise.
Such suggestions 
mirror Buçinca et al.'s~\cite{disruptive_design} recommendations to incorporate disruptive design elements that force a user to explain their 
own professional logic, prior to seeing automated content.
By forcing recruiters to explain \edit{why they have a gut feeling about a candidate or process}, this helps to create a reflection-in-action moment, moving from simply checking over genAI work to an independent evaluator.
\edit{Should such an intervention be incorporated into a workflow, fostering an agency maintenance evaluation metric---such as explaining why certain candidates are a good fit and rewarding meaningful checks---could lead to incentivizing more intentional use of AI assistance, and reduce the long-term costs of poor candidate matching.}

\paragraphnew{Outstanding Research Directions.}
Several papers have pointed to the fact that genAI systems have fundamentally changed the way that bias testing and mitigation~\cite{li2023survey} functions in AI systems, and call into question whether it is possible~\cite{anthis2025impossibility}. First, our work shows an empirical observation of a more technical point made by many~\cite{effective_testing,xiang2024fairness}: How does one mitigate bias in a textual output with an unclear mapping to the final decision? In the case of recruitment, how does one understand the possibility of AI bias in a generated rubric, or set of interview questions? More generally, as researchers, this points to a growing need to build tools to understand the impacts of intermediate model outputs in a decision-making process, not just those providing a concrete recommendation. Our work also highlights a sobering potential outcome: AI bias is not only becoming harder to find because of the structure of its output, but due to incremental over-reliance, practitioners may be less and less equipped to meaningfully watch for signs of bias (Section~\ref{sec:findings:deskilling}). This leads to a natural question: How can we help practitioners retain, or achieve, awareness of AI's influence as its role becomes more and more subtle and intertwined in every step of the human thought process towards decision-making? 

\section*{Generative AI Usage Statement}

The authors of this paper did not use any generative AI assistance when writing this manuscript.

\begin{acks}
\edit{First, the authors would like to thank their participants for sharing their time and insights. Additionally, they thank Stevie Chancellor and the reviewers for their generous feedback, which helped to improve this final manuscript.}
\end{acks}

\section*{Positionality Statement}
\edit{The authors acknowledge their experiences as researchers in relation to the topic of recruitment. As a researcher within the field of Security \& Privacy, Rosanna Bellini has experience in analyzing public tools and the effects of their usage. Emily Black, whose background is in Responsible AI, has experience in auditing AI tools, as well as analyzing the requirements around lawful use of such tools within the US regulatory framework. 
Though the authors have had minimal exposure to industry recruitment themselves, this work focuses on and presents the experience of the participants, who are all recruiting or HR professionals.}
\bibliographystyle{ACM-Reference-Format}
\bibliography{bibliography}

\appendix
\section{Taxonomy of AI-Assisted Recruitment Software}
\label{app:taxonomy}
\begin{table*}[t]
\tiny
\begin{tabularx}{0.66\columnwidth}{lXXXXXXXXXXXXXX}
& \rotatebox{45}{\textbf{\textit{Job description generation}}} & \rotatebox{45}{\textbf{\textit{Database filtering}}}    & \rotatebox{45}{\textbf{\textit{Candidate summary}}}     & \rotatebox{45}{\textbf{\textit{AI-led screening}}}      & \rotatebox{45}{\textbf{\textit{Candidate filtering}}}   & \rotatebox{45}{\textbf{\textit{Qualification grouping}}} & \rotatebox{45}{\textbf{\textit{Candidate fraud detection }}} &\rotatebox{45}{\textbf{\textit{Match ranking/scores}}}  & \rotatebox{45}{\textbf{\textit{Interview summary}}}     & \rotatebox{45}{\textbf{\textit{Interview question generation  }}} & \rotatebox{45}{\textbf{\textit{Interview transcription}}} & \rotatebox{45}{\textbf{\textit{Interview scoring}}}     & \rotatebox{45}{\textbf{\textit{Interview feedback}}} & \rotatebox{45}{\textbf{\textit{AI-led interview}}}   \\ 
\multicolumn{15}{l}{\textbf{\textit{Applicant Tracking Systems}}}\\ \hline

Workable \cite{workable}         & \multicolumn{1}{c|}{\CIRCLE} & \multicolumn{1}{c|}{\CIRCLE} & \multicolumn{1}{c|}{\CIRCLE} & \multicolumn{1}{c|}{} & \multicolumn{1}{c|}{} & \multicolumn{1}{c|}{}  & \multicolumn{1}{c|}{} & \multicolumn{1}{c|}{\CIRCLE} & \multicolumn{1}{c|}{}         & \multicolumn{1}{c|}{\CIRCLE}   & \multicolumn{1}{c|}{} & \multicolumn{1}{c|}{} & \multicolumn{1}{c|}{} & \multicolumn{1}{c|}{} \\ \hline

Ashby$^{\dag}$ \cite{ashby}           & \multicolumn{1}{c|}{\CIRCLE
} & \multicolumn{1}{c|}{\CIRCLE} & \multicolumn{1}{c|}{} & \multicolumn{1}{c|}{} & \multicolumn{1}{c|}{\CIRCLE} & \multicolumn{1}{c|}{}  & \multicolumn{1}{c|}{\CIRCLE} & \multicolumn{1}{c|}{} & \multicolumn{1}{c|}{}         & \multicolumn{1}{c|}{}   & \multicolumn{1}{c|}{\CIRCLE} & \multicolumn{1}{c|}{} & \multicolumn{1}{c|}{} & \multicolumn{1}{c|}{}\\ \hline
Gem$^{\dag*}$ \cite{gem}                             & \multicolumn{1}{c|}{} & \multicolumn{1}{c|}{\CIRCLE} & \multicolumn{1}{c|}{} & \multicolumn{1}{c|}{} & \multicolumn{1}{c|}{\CIRCLE} & \multicolumn{1}{c|}{}  & \multicolumn{1}{c|}{} & \multicolumn{1}{c|}{\CIRCLE} & \multicolumn{1}{c|}{\CIRCLE}         & \multicolumn{1}{c|}{}   & \multicolumn{1}{c|}{} & \multicolumn{1}{c|}{} & \multicolumn{1}{c|}{} & \multicolumn{1}{c|}{}\\ \hline

Breezy$^{\dag}$ \cite{breezy}           & \multicolumn{1}{c|}{}   & \multicolumn{1}{c|}{} & \multicolumn{1}{c|}{} & \multicolumn{1}{c|}{} & \multicolumn{1}{c|}{} & \multicolumn{1}{c|}{}  & \multicolumn{1}{c|}{\CIRCLE} & \multicolumn{1}{c|}{\CIRCLE} & \multicolumn{1}{c|}{\CIRCLE}         & \multicolumn{1}{c|}{}   & \multicolumn{1}{c|}{} & \multicolumn{1}{c|}{} & \multicolumn{1}{c|}{} & \multicolumn{1}{c|}{} \\ \hline
Bullhorn$^{\dag}$ \cite{bullhorn}         & \multicolumn{1}{c|}{\CIRCLE}  & \multicolumn{1}{c|}{} & \multicolumn{1}{c|}{\CIRCLE} & \multicolumn{1}{c|}{\CIRCLE} & \multicolumn{1}{c|}{} & \multicolumn{1}{c|}{}  & \multicolumn{1}{c|}{} & \multicolumn{1}{c|}{\CIRCLE} & \multicolumn{1}{c|}{\CIRCLE}         & \multicolumn{1}{c|}{\CIRCLE}   & \multicolumn{1}{c|}{} & \multicolumn{1}{c|}{} & \multicolumn{1}{c|}{} & \multicolumn{1}{c|}{} \\ \hline

iCIMS$^{\dag *}$ \cite{icims}            & \multicolumn{1}{c|}{\CIRCLE} & \multicolumn{1}{c|}{\CIRCLE} & \multicolumn{1}{c|}{} & \multicolumn{1}{c|}{} & \multicolumn{1}{c|}{} & \multicolumn{1}{c|}{}  & \multicolumn{1}{c|}{} & \multicolumn{1}{c|}{\CIRCLE} & \multicolumn{1}{c|}{}         & \multicolumn{1}{c|}{\CIRCLE}   & \multicolumn{1}{c|}{} & \multicolumn{1}{c|}{} & \multicolumn{1}{c|}{} & \multicolumn{1}{c|}{} \\ \hline

Greenhouse$^{\dag}$ \cite{greenhouse}                           & \multicolumn{1}{c|}{\CIRCLE}   & \multicolumn{1}{c|}{} & \multicolumn{1}{c|}{} & \multicolumn{1}{c|}{} & \multicolumn{1}{c|}{\CIRCLE} & \multicolumn{1}{c|}{}  & \multicolumn{1}{c|}{} & \multicolumn{1}{c|}{} & \multicolumn{1}{c|}{\CIRCLE}         & \multicolumn{1}{c|}{\CIRCLE}   & \multicolumn{1}{c|}{} & \multicolumn{1}{c|}{} & \multicolumn{1}{c|}{} & \multicolumn{1}{c|}{} \\ \hline

Workday$^{\dag}$ \cite{workday}                           & \multicolumn{1}{c|}{\CIRCLE}   & \multicolumn{1}{c|}{\CIRCLE} & \multicolumn{1}{c|}{} & \multicolumn{1}{c|}{\CIRCLE} & \multicolumn{1}{c|}{} & \multicolumn{1}{c|}{\CIRCLE}  & \multicolumn{1}{c|}{} & \multicolumn{1}{c|}{} & \multicolumn{1}{c|}{}         & \multicolumn{1}{c|}{}   & \multicolumn{1}{c|}{} & \multicolumn{1}{c|}{} & \multicolumn{1}{c|}{} & \multicolumn{1}{c|}{} \\ \hline

\multicolumn{15}{l}{\textbf{\textit{Supplemental Tooling}}}\\ \hline

Indeed$^{\dag}$ \cite{indeed}         & \multicolumn{1}{c|}{\CIRCLE}   & \multicolumn{1}{c|}{\CIRCLE} & \multicolumn{1}{c|}{\CIRCLE} & \multicolumn{1}{c|}{} & \multicolumn{1}{c|}{} & \multicolumn{1}{c|}{}  & \multicolumn{1}{c|}{} & \multicolumn{1}{c|}{\CIRCLE} & \multicolumn{1}{c|}{}         & \multicolumn{1}{c|}{}   & \multicolumn{1}{c|}{} & \multicolumn{1}{c|}{} & \multicolumn{1}{c|}{} & \multicolumn{1}{c|}{} \\ \hline

LinkedIn Recruiter$^{\dag *}$ \cite{linkedinrecruiter}     & \multicolumn{1}{c|}{}   & \multicolumn{1}{c|}{\CIRCLE} & \multicolumn{1}{c|}{\CIRCLE} & \multicolumn{1}{c|}{} & \multicolumn{1}{c|}{} & \multicolumn{1}{c|}{\CIRCLE}  & \multicolumn{1}{c|}{} & \multicolumn{1}{c|}{} & \multicolumn{1}{c|}{}         & \multicolumn{1}{c|}{}   & \multicolumn{1}{c|}{} & \multicolumn{1}{c|}{} & \multicolumn{1}{c|}{} & \multicolumn{1}{c|}{} \\ \hline
Metaview$^{\dag *}$ \cite{metaview}       & \multicolumn{1}{c|}{\CIRCLE}   & \multicolumn{1}{c|}{\CIRCLE} & \multicolumn{1}{c|}{} & \multicolumn{1}{c|}{} & \multicolumn{1}{c|}{} & \multicolumn{1}{c|}{}  & \multicolumn{1}{c|}{} & \multicolumn{1}{c|}{} & \multicolumn{1}{c|}{\CIRCLE}         & \multicolumn{1}{c|}{}   & \multicolumn{1}{c|}{\CIRCLE} & \multicolumn{1}{c|}{} & \multicolumn{1}{c|}{} & \multicolumn{1}{c|}{} \\ \hline
Hirevue$^{\dag}$ \cite{hirevue}        & \multicolumn{1}{c|}{}  & \multicolumn{1}{c|}{} & \multicolumn{1}{c|}{} & \multicolumn{1}{c|}{} & \multicolumn{1}{c|}{} & \multicolumn{1}{c|}{}  & \multicolumn{1}{c|}{} & \multicolumn{1}{c|}{} & \multicolumn{1}{c|}{\CIRCLE}         & \multicolumn{1}{c|}{\CIRCLE}   & \multicolumn{1}{c|}{\CIRCLE} & \multicolumn{1}{c|}{\CIRCLE} & \multicolumn{1}{c|}{} & \multicolumn{1}{c|}{} \\ \hline
BrightHire$^{\dag}$ \cite{brighthire} & \multicolumn{1}{c|}{\CIRCLE}  & \multicolumn{1}{c|}{} & \multicolumn{1}{c|}{} & \multicolumn{1}{c|}{\CIRCLE} & \multicolumn{1}{c|}{} & \multicolumn{1}{c|}{}  & \multicolumn{1}{c|}{} & \multicolumn{1}{c|}{} & \multicolumn{1}{c|}{\CIRCLE}         & \multicolumn{1}{c|}{\CIRCLE}   & \multicolumn{1}{c|}{} & \multicolumn{1}{c|}{} & \multicolumn{1}{c|}{\CIRCLE} & \multicolumn{1}{c|}{} \\ \hline
Willo$^{\dag}$ \cite{willo}            & \multicolumn{1}{c|}{}    & \multicolumn{1}{c|}{} & \multicolumn{1}{c|}{} & \multicolumn{1}{c|}{} & \multicolumn{1}{c|}{\CIRCLE} & \multicolumn{1}{c|}{}  & \multicolumn{1}{c|}{} & \multicolumn{1}{c|}{} & \multicolumn{1}{c|}{\CIRCLE}         & \multicolumn{1}{c|}{\CIRCLE}   & \multicolumn{1}{c|}{\CIRCLE} & \multicolumn{1}{c|}{\CIRCLE} & \multicolumn{1}{c|}{} & \multicolumn{1}{c|}{} \\ \hline

Juicebox$^{\dag *}$ \cite{juicebox_juicebox_nodate}            & \multicolumn{1}{c|}{}   & \multicolumn{1}{c|}{\CIRCLE} & \multicolumn{1}{c|}{\CIRCLE} & \multicolumn{1}{c|}{} & \multicolumn{1}{c|}{} & \multicolumn{1}{c|}{}  & \multicolumn{1}{c|}{} & \multicolumn{1}{c|}{} & \multicolumn{1}{c|}{}         & \multicolumn{1}{c|}{}   & \multicolumn{1}{c|}{} & \multicolumn{1}{c|}{} & \multicolumn{1}{c|}{} & \multicolumn{1}{c|}{} \\ \hline

\multicolumn{15}{l}{\textbf{\textit{Agentic Software}}}\\ \hline

Eightfold \cite{eightfold} & \multicolumn{1}{c|}{\CIRCLE} & \multicolumn{1}{c|}{\CIRCLE} & \multicolumn{1}{c|}{} & \multicolumn{1}{c|}{\CIRCLE} & \multicolumn{1}{c|}{} & \multicolumn{1}{c|}{}  & \multicolumn{1}{c|}{} & \multicolumn{1}{c|}{\CIRCLE} & \multicolumn{1}{c|}{\CIRCLE}         & \multicolumn{1}{c|}{}   & \multicolumn{1}{c|}{\CIRCLE} & \multicolumn{1}{c|}{\CIRCLE} & \multicolumn{1}{c|}{} & \multicolumn{1}{c|}{\CIRCLE} \\ \hline

Covey \cite{covey}            & \multicolumn{1}{c|}{}   & \multicolumn{1}{c|}{\CIRCLE} & \multicolumn{1}{c|}{} & \multicolumn{1}{c|}{} & \multicolumn{1}{c|}{} & \multicolumn{1}{c|}{}  & \multicolumn{1}{c|}{\CIRCLE} & \multicolumn{1}{c|}{\CIRCLE} & \multicolumn{1}{c|}{}         & \multicolumn{1}{c|}{}   & \multicolumn{1}{c|}{} & \multicolumn{1}{c|}{} & \multicolumn{1}{c|}{} & \multicolumn{1}{c|}{} \\ \hline

\end{tabularx}

\caption{Software is organized the AI functionalities it supports and ordered by recruitment stage (Section~\ref{sec:backlit}). Applicant tracking systems refer to platforms that support the full-cycle recruitment pipeline, while supplemental tooling offers tooling that focus on a single pipeline phase. Agentic software or agents refer to the distinct functionality of autonomous action that can operate with minimal human intervention (identified by ${^*}$). For clarity, we denote software that participants reported to actively use with $^{\dag}$}.
\Description{A table containing the recruitment-specific tools mentioned in participant interviews and their AI features. The most common features include job description generation, database filtering, and summarizing interviews. Most software providers only support a handful of AI features, where the provider with the most features supports eight out of 14 AI features.}
\label{tab:taxonomy_vis}
\vspace{-1cm}
\end{table*}

In parallel to our interview analyses, we performed an analysis of commonly discussed HR-specific hiring tools. Our taxonomy can be found in Table~\ref{tab:taxonomy_vis}. Through this analysis, we found that there exist AI, and specifically genAI, applications throughout the first four stages of the recruitment process (Section~\ref{sec:backlit}), excluding the extension of the final offer. 
The most common application of AI is `Database filtering,' which facilitates sourcing by searching across candidate databases, though this functionality existed prior to 2022 and the introduction of ChatGPT. `Job Description Generation,' which requires genAI technology, is the second most prevalent application of AI. The stage with the most genAI-specific applications is `Interviewing,' where genAI has introduced new functionalities, such as summarizing interviews and generating interview questions. The AI functionalities that participants did not explicitly mention were used include `AI-led screening,' `Candidate filtering,' `Interview scoring,' `Interview feedback,' and `AI-led interview.'

\section{Interview Protocol}
\label{app:protocol}
We provide our complete semi-structured interview protocol below. Due to the nature of semi-structured interviews, we asked follow-up questions and omitted other questions in the protocol as needed.

\subsection*{Grounding Questions}
\noindent \textbf{Q1} Tell me about your work and your organization.
\begin{enumerate}
    \itemsep0em
    \item How long have you been in this role?
\end{enumerate}

\noindent \textbf{Q2} Can you tell me about how you are involved with company hiring processes, and a bit about what that process looks like (think of a recent hire)?
\begin{enumerate}
    \itemsep0em
    \item Are there steps you normally go through? Frequency? Number of people involved?
    \item Is there a rubric you use to evaluate people? What factors do you often consider?
\end{enumerate}

\subsection*{General Use of GenAI in Hiring}
\noindent \textbf{Q1} Have you used an LLM to support you with your work related to hiring? If so, how so and where in the process do you use it?
\begin{enumerate}
    \itemsep0em
    \item ChatGPT, Claude are popular tools, what model do you use?
    \item Is this a general model or personalised to your company?
    \item Is this a personal or business account?
    \item What motivated you to use it?
\end{enumerate}

\noindent \textbf{Q2} What do you ask the LLM to do <in the hiring process stated previously>?
\begin{enumerate}
    \itemsep0em
    \item Are there specific tasks you assign to this?
    \item What information do you give it?
    \item What do you do with the output? \begin{enumerate}
        \itemsep 0em
        \item Is this useful to you? If yes/no, why?
    \end{enumerate}
\end{enumerate}

\subsection*{Honing in on One Application of GenAI}
\noindent \textbf{Q1} What are the inputs/outputs at this part of the process?
\begin{enumerate}
    \itemsep0em
    \item How do you use the outputs to reach the next stage of the decision process?
\end{enumerate}
\noindent \textbf{Q2} Has your use of AI matched with your expectations?

\noindent \textbf{Q3} You mention <AI-assisted task>---have you ever looked at the <AI output, e.g. summary of an interview> and felt that it changed your impression of <the input, e.g. a candidate/the position>?
\begin{enumerate}
    \itemsep0em
    \item Has it brought things to your attention you wouldn’t have noticed on your own? 
\end{enumerate}
\noindent \textbf{Q4} Do you find yourself deferring to the <AI output>?
\begin{enumerate}
    \itemsep0em
    \item Do you think you do so just because it’s convenient?
    \item Do you think these tools are improving the selection process?
    \item What factors make you trust or question its judgment even when your professional instinct disagrees?
\end{enumerate}
\noindent \textbf{Q5} Do you ever check for errors on the outputs of the LLM? What do you look for?

\subsection*{Perception of Agency over Choices}
\noindent \textbf{Q1} Is there a reason you haven’t used LLMs in other parts of your hiring process? 
\begin{enumerate}
    \itemsep0em
    \item Is there something you would never ask an LLM to do for you? 
    \item Do you have any concerns about how it is currently used or could be used in your company?
\end{enumerate}
\noindent \textbf{Q2} Do you feel like you understand why it produces the candidates, scores, summaries etc. that it does?
\begin{enumerate}
    \itemsep 0em
    \item Has there been anything that challenges this perspective?
\end{enumerate}
\noindent \textbf{Q3} How much do you feel like you are still in control of the final decision, versus the system guiding or constraining your choices?

\noindent \textbf{Q4} Do you feel like you’ve lost any skills as a result of relying on AI? 
\begin{enumerate}
    \itemsep 0em
    \item If the AI suddenly stopped working tomorrow, would you know how to tackle something?
\end{enumerate}

\noindent \textbf{Q5} How do you explain your hiring choices to others (e.g., managers) when the AI system has played a role in shaping those choices?

\noindent \textbf{Q6} <If participant uses genAI in applicant evaluation> Can you give me an example of a prompt you’ve used to evaluate applicants with an LLM? This prompt can be generic.

\section{Codebook}
\label{app:codebook}
See Table~\ref{fig:codebook} for the codebook resulting from our process of thematic analysis.

\begin{table*}[]
\small
\begin{tabularx}{\linewidth}{|X|X|}
\hline
\textbf{Company-focused} & \textbf{Candidate-focused} \\ \hline
{\begin{tabularx}{\linewidth}[t]{@{}X@{}}
\textit{Business Culture}: What is the company's culture like?\\ 
\textit{Business Size}: How large is the company\\
\textit{Pressure to Use AI in Business}: Participant reporting pressure to use AI from their employer\\ 
\textit{Stage of AI Adoption}: How much AI the employer is incorporating company-wide\\ 
\textit{Prohibitive Cost of AI}: Statement that AI tools are out of the price range for a participant's company\end{tabularx}} 
& 
 {\begin{tabularx}{\linewidth}[t]{@{}X@{}}
\textit{Candidate Use of AI}: How candidates use AI\\ 
\textit{Candidate Effort stands out}: The manual effort of a candidate is obvious when reviewing their application\\ 
\textit{Desired Candidate Skillsets}: Skills the participant looks for\\ 
\textit{Applicant-recruiter Arms Race}: Participant's use of AI to combat candidate's use\\ 
\textit{Fake Candidates}: Discussion of reportedly ``fake'' candidates\\ 
\textit{High Volume of Applications}: Capturing the great demand on recruiters for reviewing applications\end{tabularx}} \\ \hline
\textbf{Agency-focused} & \textbf{Impacts/Future}  \\ \hline
{\begin{tabularx}{\linewidth}[t]{@{}X@{}}
\textit{Auto/AI-Resistant Work}: Assertion that a specific task cannot be performed by AI\\ 
\textit{Deference to AI}: Instance where the participant trusts AI\\ 
\textit{Employer AI Red Line}: Determination of when AI should not be used in the recruitment process\\ 
\textit{Human Agency Maintained}: Belief that the participant maintains decision-making agency\\ 
\textit{Lack of Trust in AI/Skepticism}: Skepticism toward the use of AI in general or a specific task\\ 
\textit{Shared Decision-making}: Belief that AI has some impact on the decision-making process\\ 
\textit{Perceived Human Manual}: Claim that participant is manually performing a certain task\end{tabularx}} &

{\begin{tabularx}{\linewidth}[t]{@{}X@{}}
\textit{AI Future Potential}: Looking forward to AI's future impact\\ 
\textit{Legal and Privacy Risk}: Reported fear of risk or performing actions that constitute risk\\ 
\textit{Loss of Skill/Deskilling}: Observed behavior that reflects deskilling\\ 
\textit{Overpromising of AI}: Belief that AI marketing has overpromised on the utility of AI\\ 
\textit{AI Used Differently by Generations}: Pattern that different age groups use AI in disparate manners\\ 
\textit{Two-tiered AI use}: Different usage of AI depending on qualities of the role or candidate\end{tabularx}}\\ \hline

\textbf{Participant/Role Information} & \textbf{Recruitment Stage}\\ \hline

{\begin{tabularx}{\linewidth}[t]{@{}X@{}}
\textit{Job Level Type}: The level of position that the participant hires\\ 
\textit{Participant Background}: Information about participant's role and experience\\ 
\textit{Self-Assessment of AI Proficiency}: A participant's confidence in their use of AI\\ 
\textit{Time in Role}: How long has the participant been in this role\end{tabularx}} & 
   {\begin{tabularx}{\linewidth}[t]{@{}X@{}}
\textit{Assessment}: Process of forming a judgment on a candidate\\ \textit{Conceptualizing/Themeing}: Generation of a job description\\ \textit{Final Decision-making}: Selection of candidate(s) to send an~offer\\ 
\textit{Interviewing}: Conversations with candidates\\ 
\textit{Logistics}: Planning interviews, sending communication, etc.\\ 
\textit{Sourcing}: Finding candidates to apply to a job opening\\ 
\textit{Screening}: Comparing a candidate's skills and qualifications to the job requirements\end{tabularx}}\\ \hline
\textbf{Task Type}                                             & \textbf{Performance of AI Systems} \\ \hline
{\begin{tabularx}{\linewidth}[t]{@{}X@{}}
\textit{Agentic AI}: Using AI agents to complete a task\\ 
\textit{AI Content Generation}: Using genAI to produce content\\ 
\textit{Analogue-Technical Processes}: A process combining both technical and manual methods\\ 
\textit{Grouping}: Classifying candidates in a fixed number of groups\\ 
\textit{Keyword Searching}: Looking for specific keywords \\ 
\textit{Manual Work}: Human-only completion of a recruitment task\\ 
\textit{Transcription}: Translation from a recording to text\\ 
\textit{Summarizing}: Reduction of an input into a shortened form\\ 
\textit{Ranking}: Comparing candidates to one another individually\\ 
\textit{Research/Information Gathering}: Process of learning about a certain domain\\ 
\textit{System Indicators of Job Compatibility}: How is job compatibility reflected to the participant\end{tabularx}}
& {\begin{tabularx}{\linewidth}[t]{@{}X@{}}
\textit{AI Evaluates Candidates Badly}: Inconsistent or incorrect evaluation of candidates\\ 
\textit{AI is Biased}: Relaying of concerns about biased outputs from~AI \\ 
\textit{AI Lacks Nuance}: Discusses that AI systems may be brittle or do not understand humans\\ 
\textit{AI Makes More More Work}: Claim that the participant's use of AI increases their workload\\ 
\textit{AI Reveals Overlooked Details}: AI provides the participant with a more detailed interpretation of the input\\ 
AI Saves Time: The use of AI reduces time spent on a task\\ 
\textit{AI Tool Performance}: Reflects how well the AI tool is seen to perform in general\\ 
\textit{AI Use is Essential for Efficiency}: Belief that one's use of AI is essential for the efficient completion of tasks\end{tabularx}}                                                                                                                                         \\ \hline
\end{tabularx}
\caption{The codebook derived from our thematic analysis, grouped by the codes' focus.}
\Description{A table showing the final codes from our reflexive thematic analysis. The codes cover eight different foci: the company, the candidates, participants' agency, the impacts of AI use and the future use of AI, participant/role information, the stage of recruitment being discussed, the type of task being discussed, and how well the AI system performs.}
\label{fig:codebook}
\end{table*}

\end{document}